\documentclass[rnote,twocolumn,traditabstract]{aa} 
\usepackage{epsfig}
\usepackage{graphicx}
\usepackage{times}
\usepackage{natbib}

\bibpunct{(}{)}{;}{a}{}{,} 

\def\5{\sc v}
\def\4{\sc iv}
\def\3{\sc iii}
\def\2{\sc ii}
\def\1{\sc i}
\def\lam{$\lambda$}

\def\o{$\phantom{1}$}

\begin{document}

\title{A FEROS spectroscopic study of the extreme O supergiant 
He\,3--759\thanks{Based on observations made with ESO telescopes at the 
La Silla observatory under program ID 082-D.0136}
}

\author{P.~A.~Crowther\inst{1} \& C.~J.~Evans\inst{2}}
\offprints{Paul.Crowther@sheffield.ac.uk}
\authorrunning{P.~A.~Crowther \& C.~J.~Evans}
\titlerunning{FEROS spectroscopy of He\,3--759}
\institute{
           Department of Physics \& Astronomy, 
           Hicks Building, 
           University of Sheffield, 
           Hounsfield Road, 
           Sheffield, S3 7RH, UK
           \and
           UK Astronomy Technology Centre, 
           Royal Observatory Edinburgh, 
           Blackford Hill, Edinburgh, EH9 3HJ, UK
           }
\date{}

\abstract{We present a study of the extreme O-type supergiant He\,3--759 using 
new high-resolution FEROS data, revealing that it is a near 
spectroscopic twin of HD~151804 (O8\,Iaf). We investigate the extinction 
towards He\,3--759 using a variety of methods, revealing $A_{\rm V} \sim 
4.7^{\rm m}$. If we assume He\,3--759 has an identical absolute $K$-band
magnitude to HD\,151804 we find that it lies in the Sagittarius-Carina 
spiral arm at a distance of $\sim$6.5 kpc. We derive the physical and wind 
properties for He\,3--759, revealing $T_{\ast}$ = 30.5 kK, $\log 
L/L_{\odot}$ = 5.9 and $\dot{M} = 10^{-5.17}$ $M_{\odot}$\,yr$^{-1}$ for 
a clumped wind
whose terminal velocity is estimated at 1000 km\,s$^{-1}$. The atmosphere of He\,3--759 
is  enriched in helium  ($X_{\rm He}$ = 49\%) and nitrogen 
($X_{\rm N}$ = 0.3\%). A reanalysis of HD\,151804 and HD\,152408 (WN9ha) reveals similar 
parameters except that the WN9ha star possesses a stronger wind and 
reduced surface hydrogen content. HD\,151804 and HD\,152408 lie within the 
Sco~OB1 association, with initial masses of $\sim 60\,M_{\odot}$ and ages 
$\sim2.7$ Myr, consistent with NGC\,6231 cluster members using standard
Geneva isochrones.  Improved agreement with observed surface 
abundances are obtained for similar initial masses with more  recent Geneva group predictions 
from which higher ages of $\sim$3.75 Myr are obtained. No young, 
massive star cluster is known to be associated with He\,3--759.}


\maketitle 

\section{Introduction}


In normal star-forming galaxies, massive O-type stars dominate both the 
Lyman continuum ionizing budget and the feedback of mechanical energy 
through their intense stellar winds and, ultimately, as core-collapse 
supernovae. The bulk of their short (3--10\,Myr) lives are spent on the 
main sequence as an unevolved dwarf or giant, before rapidly shedding 
their hydrogen envelope during either the Red Supergiant, Luminous Blue 
Variable or Wolf-Rayet phase. O-type supergiants represent the transition 
between these stages for the highest mass stars, with characteristic 
emission lines of He\,{\sc ii} \lam4686 and H$\alpha$ due to their 
relatively strong stellar winds. N\,{\sc iii} \lam\lam4634-41 emission
is also commonly seen in such stars, with a corresponding Of nomenclature.

Surveys of emission-line O-type stars in the Small and Large Magellanic 
Clouds are relatively complete due to the low interstellar extinction 
towards their sight-lines. Comparable Milky Way surveys remain highly 
incomplete with the exception of known OB associations in the solar 
neighbourhood \citep[e.g.][]{h78}, although H$\alpha$ surveys such as 
IPHAS \citep{iphas, iphas_2} and VPHAS+ \citep{vphas} are in the process 
of remedying this deficit, at least for sources detected optically.

Still, many sources from the extensive Michigan-Mt Wilson southern 
H$\alpha$ survey \citep{h76} remain largely neglected. He\,3--759 is one 
such source, and is the focus of the present study. This was first 
reported in the catalogue of Galactic Wolf-Rayet (WR) stars by 
\citet[][Star~\#41] {r62} due to broad H$\alpha$ emission. 
The intensity and sharpness of H$\beta$, He\,{\sc ii} \lam4686 and N\,{\sc 
iii} \lam\lam4634-41 emission led \citet{ch79} to favour an Of 
classification. However, ultraviolet spectroscopy from \citet{s90} 
suggested a contradictory early B-type classification, providing the 
motivation for the present study. Subsequently, \citet{d01} included 
He\,3--759 in their photometric catalogue of southern emission line 
sources, assigning it a (classical) Be spectral type \citep[see 
also][]{t94}.


New observations are reported in Sect.~\ref{SECT2}, with estimates of the 
distance and extinction towards He\,3--759 obtained in Sect.~\ref{SECT3}. 
A spectroscopic analysis is performed in Sect.~\ref{SECT4}, including a 
comparison to HD~151804 (O8\,Iaf) and HD\,152408 (WN9ha), and conclusions 
are drawn in Sect.~\ref{SECT5}.

\begin{table}
\caption[]{Published coordinates of He\,3--759, including
astrometry from Tycho-2.}\label{astrometry}
\begin{center}
\begin{tabular}{lll}
\hline
$\alpha$ (J2000) & $\delta$ (J2000) & Reference\\
\hline
12 11.3 & --62 29 & \citet{r62}; \citet{h76} \\
12 11.3 & --62 30 & \citet{ch79} \\
12 12 08.56 & --62 29 00.6 & \citet{t94}\\
12 12 08.6 & --62 29 01 & \citet{d01} \\
12 11 18.54 & --62 29 43.5 & Tycho-2 \\
\hline
\end{tabular}
\end{center}
\end{table}

\section{Observations}\label{SECT2}

Previously unpublished spectroscopy of He\,3--759 was obtained with 
the Double Beam Spectrograph (DBS) mounted at the Australian National 
University (ANU) 2.3m telescope in April 1996.  Subsequent high-resolution 
spectroscopy was obtained with the Fibre-fed Extended Range Optical 
Spectrograph (FEROS) at the 2.2-m Max Planck Gesellschaft (MPG)/European 
Southern Observatory (ESO) telescope in March 2009. 

\begin{table*}
\caption[]{Visual (Tycho-2 B$_{\rm T}$ and V$_{\rm T}$ in parenthesis) 
and near-IR photometry for the O8\,Iaf stars He\,3--759 and HD\,151804, including a distance 
estimate to He\,3--759.}
\label{photom}
\begin{center}
\begin{tabular}{lcclllllllllll}
\hline
Star          & V & B-V & K$_{s}$ & J--K$_{s}$ & H--K$_{s}$ & (J--K$_{s}$)$_0$ & (H--K$_{s}$)$_0$ & $A_{\rm K_{s}}^{J-K}$ & $A_{\rm K_{s}}^{H-K}$ & $A_{\rm K_{s}}$ & DM & & $M_{\rm 
K_{s}}$ \\
\hline
He\,3--759    &(11.45)&(1.01) &7.88 & 0.66 & 0.32 &--0.09 &--0.04 &0.50 &0.65 &0.58 & 14.0 & $\leftarrow$ & 
--6.7 \\
HD~151804     &\o5.22&0.07&4.86$^{\ddag}$& 0.20$^{\ddag}$ & 
                               0.06$^{\ddag}$ & --0.09 &--0.04 &0.19 &0.18 &0.19 & 11.4 & $\rightarrow$ & --6.7 \\
\hline
\end{tabular}
\note{$\ddag$ JHK magnitudes of \citet{lw84} are preferred to
2MASS due to a low quality index in this instance.}
\end{center}
\end{table*}

\subsection{Coordinates of He\,3--759}

The new observations highlighted discrepancies in previously published 
coordinates of He\,3--759.  These are summarised in 
Table~\ref{astrometry}, with coordinates precessed to J2000 epochs using 
the {\sc starlink coco} package where necessary.  Note that the 
coordinates listed by the {\sc simbad} 
database\footnote{http://simbad.u-strasbg.fr/simbad/} from \citet{t94} are 
incorrect. The positions published by \citet{d01} are presumably rounded 
values from \citeauthor{t94}

FEROS observations of He\,3--759 were initially attempted on 2009 March 18 
using the {\sc simbad} coordinates. However the resulting spectrum was, 
surprisingly, of a cool M-type star, namely the long period variable IRAS 
12094--6212 from \citet{c91}. We subsequently inspected the fits header 
information from the ANU/DBS spectroscopy, which were consistent with 
\citet{ch79} values, although 5\farcm45 away from the \citet{t94} 
position. Accurate astrometry of He\,3--759 from Tycho-2 is included in 
Table~\ref{astrometry}.

%


\begin{figure*}
\begin{center}
\includegraphics[width=18cm]{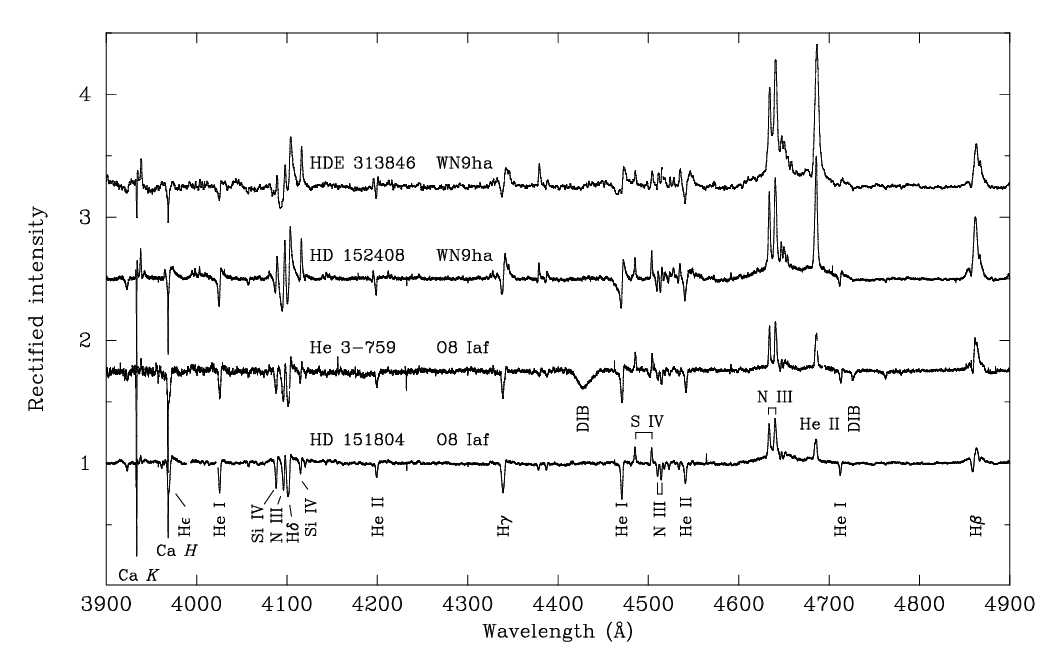}
\caption[]{Blue region FEROS spectrum of He\,3--759, compared with the AAT-UCLES 
spectra of HD\,151804, HD\,152408 and HDE\,313846 from \citet{cb97}.
Spectral lines identified in HD\,151804 are He~{\sc i}
\lam\lam4026, 4713; Si~{\sc iv} \lam\lam4089, 4116; N~{\sc iii} \lam\lam4097,
4511-15 absorption, \lam\lam4634-40-42 emission; the H$\epsilon$, H$\delta$,
H$\gamma$ and H$\beta$ Balmer lines; He~{\sc ii} \lam\lam4200, 4542
absorption, \lam4686 emission; S~{\sc iv} \lam\lam4486-4504
emission.  The \lam\lam4428, 4727 DIBs are marked in the He\,3--759
spectrum which, for clarity, has been 11-pixel median filtered.}\label{blue}

\includegraphics[width=18cm]{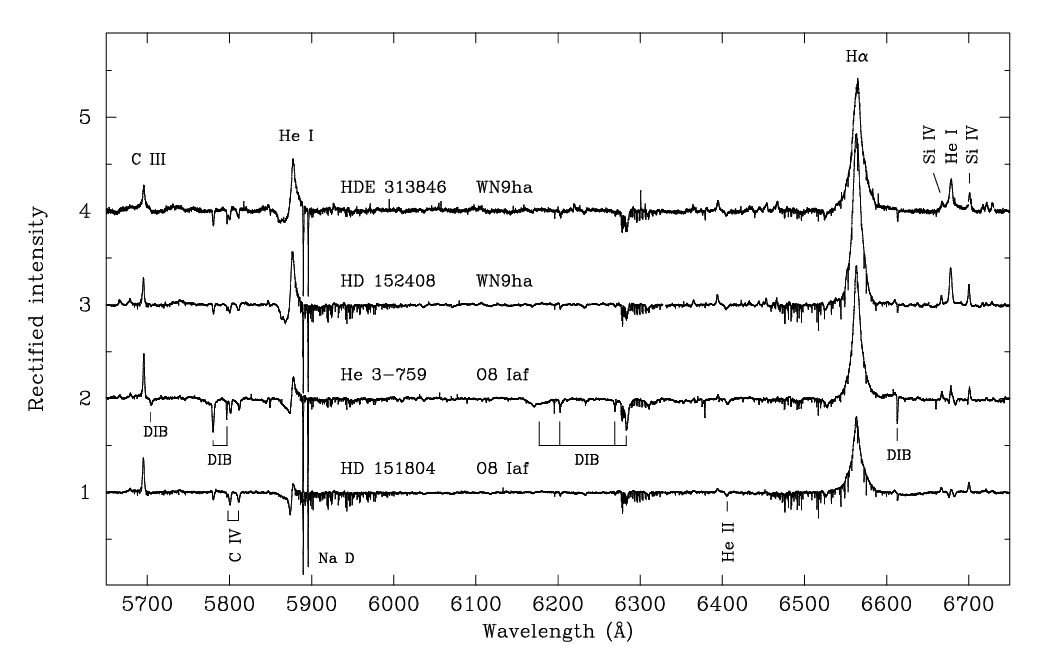}
\caption[]{Green-red region FEROS spectrum of He\,3--759, compared with the AAT-UCLES 
spectra of HD\,151804, HD\,152408 and HDE\,313846 from \citet{cb97}.
Emission lines identified in HDE\,313846 are C~{\scriptsize III} \lam5696, He~{\scriptsize I}
\lam\lam5876, 6678, H$\alpha$, and Si~{\scriptsize IV} \lam\lam6667, 6700.  Absorption lines
marked in HD\,151804 are C~{\scriptsize IV} \lam\lam5801, 5812,
He~{\scriptsize II} \lam6406, and the strong interstellar Na~D lines.
The strong DIBs marked in the He\,3--759 spectrum are
\lam\lam5705, 5780, 5797, 6177, 6203, 6269, 6283, 6613.}\label{red}
\end{center}
\end{figure*}


\subsection{ANU/DBS spectroscopy}

We used DBS at the ANU 2.3m telescope to obtain blue, yellow and
red spectroscopy of He\,3--759 on 1996 Apr 1--3. The detector
for both arms of DBS were  1752$\times$532 pix SITE CCDs with
blue and red 1200  l\,mm$^{-1}$ gratings providing a dispersion of
0.5\AA\,pix$^{-1}$. The blue DBS arm was  used on 1 Apr to obtain
1\AA\ (2 pix) resolution spectroscopy of \lam3980--4975, with 
the red DBS arm used on Apr 2 and 3 to obtain 1\AA\ resolution
spectroscopy of the \lam5695--6700 and \lam4955--5955 regions,
respectively. A standard CCD reduction was followed, enabling
He\,3--759 to be confirmed as an Of star. However, the modest S/N achieved 
($\sim$10--40) was inadequate for a quantitative analysis. A relative
flux calibration was also achieved, which was absolutely calibrated
using the Tycho-2 filter magnitudes, from which V$\sim$11.3$^{\rm m}$
is estimated. 


\subsection{MPG-ESO 2.2m/FEROS spectroscopy}

Two 1800s exposures of He\,3--759 were obtained with FEROS on the nights 
of 2009 March 19 \& 20.  FEROS is a cross-dispersed, fixed configuration 
instrument \citep{feros}, which delivers $R=48,000$ on the 2.2-m, with 
continuous spectral coverage of $\sim$3600-9200\AA.  The spectra 
presented here are from the reduction pipeline that runs at the 
telescope; subsequent checks with reduction routines tailored for FEROS 
\citep[as used by][]{s09} yielded indistinguishable final spectra.

\subsection{Spectral Classification}

The blue and green-red spectral regions of He\,3--759 are presented in
Figures~\ref{blue} and \ref{red}, respectively.  Also shown are
high-resolution spectra from \citet{cb97} of HD\,151804, HD\,152408,
and HDE\,313846 from the Anglo-Australian Telescope (AAT) using the
University College London Echelle Spectrograph (UCLES).  HD\,151804 is
a `normal' Of star, classified as O8~Iaf \citep{ca71, w72}, with the more
extreme sources HD\,152408 and HDE\,313846 (WR108) reflected by their 
classification of WN9ha \citep{cb97, bc99}.

These spectra illustrate an elegant morphological sequence in terms of 
increasing emission-line intensities.  The N\,{\sc iii}, He\,{\sc ii} and 
H$\beta$ emission in He\,3--759 is slightly stronger than in HD\,151804, 
but otherwise their blue (stellar) spectra are very similar 
(Figure~\ref{blue}), including the diagnostic
He\,{\sc i} \lam4471/He\,{\sc ii} \lam4542 ratio\footnote{We measure
$\log 4471/4542$ = 0.19 from our FEROS dataset for 
He\,3--759 versus 0.22 from our UCLES observations of HD\,151804.}
and we therefore adopt an O8~Iaf classification for 
He\,3--759. Its appearance confirms the description of \citet{ch79},
with He\,3--759 and HD\,151804 among the rare subset of O stars in which
H$\beta$ is observed in emission, signifying extreme mass-loss properties.

The green-red region (Figure~\ref{red}) reinforces the notion that
He\,3--759 is a slightly more extreme Of star than HD\,151804, with
stronger H$\alpha$ and He\,{\sc i} 5876 emission.  However, its 
emission features
are not as pronounced as in  HD\,152408 and
HDE\,313846, and a normal Of supergiant classification is sufficient.
Also, note the increasing intensities of the Si\,{\sc iv} \lam\lam6667, 
6701 and He\,{\sc i} \lam6678 emission in the sequence.

\section{Reddening \& Distance}\label{SECT3}

Reliable photometry of He\,3--759 is somewhat sparse in the literature
so we consider three different approaches to estimate its reddening
drawn from (i) infrared photometry from the Two Micron All Sky Survey
\citep[2MASS,][]{2mass}; (ii)  
ultraviolet International Ultraviolet Explorer (IUE) spectroscopy
from \citet{s90}; (iii) the strength of Diffuse Interstellar Band (DIBs)
observed in the FEROS spectroscopy. An estimate of the distance to
He\,3--759 then follows from comparison with HD\,151804 which is
a member of the Sco~OB1 association \citep[distance modulus 11.4$^{\rm m}$, ][]{h78}.
thence its distance, for quantitative analysis.

\subsection{Photometry}

A summary of visible and near-IR photometry for He\,3--759 is presented in Table~\ref{photom}.
Optical measurements are rather heterogeneous, including the Tycho-2
(V$_{\rm T}$ = 11.45), 
2nd USNO CCD (10.94) and USNO-B1.0 catalogues (B1=12.35, R1=10.85,
I=10.38). We include visual and IR photometry of HD\,151804 (O8\,Iaf) drawn from
\citet{lw84}, \citet{cb97} and references therein. Intrinsic near-IR colours
are obtained from our analysis of He\,3--759 (Sect.~\ref{SECT4}) from
which $K_{s}$-band extinctions, $A_{\rm K_{s}}$, may be obtained using the  extinction 
relations from \citet{i05}. Our derived extinction of $A_{\rm K_{s}}$ = 0.58 $\pm
0.08^{\rm m}$ for He\,3--759 
corresponds to $E_{B-V}$=1.65, assuming a standard Galactic extinction law.
A similar approach for HD~151804 reveals $A_{\rm K_{s}}$=0.19$^{\rm m}$, or 
$E_{B-V}$=0.54$^{\rm m}$, rather higher than results derived previously, 
such as
$E_{B-V}$=0.32$^{\rm m}$ from \citet{cb97}.

\subsection{Archive Ultraviolet Spectroscopy}

\begin{figure}
\begin{center}
\includegraphics[width=\columnwidth]{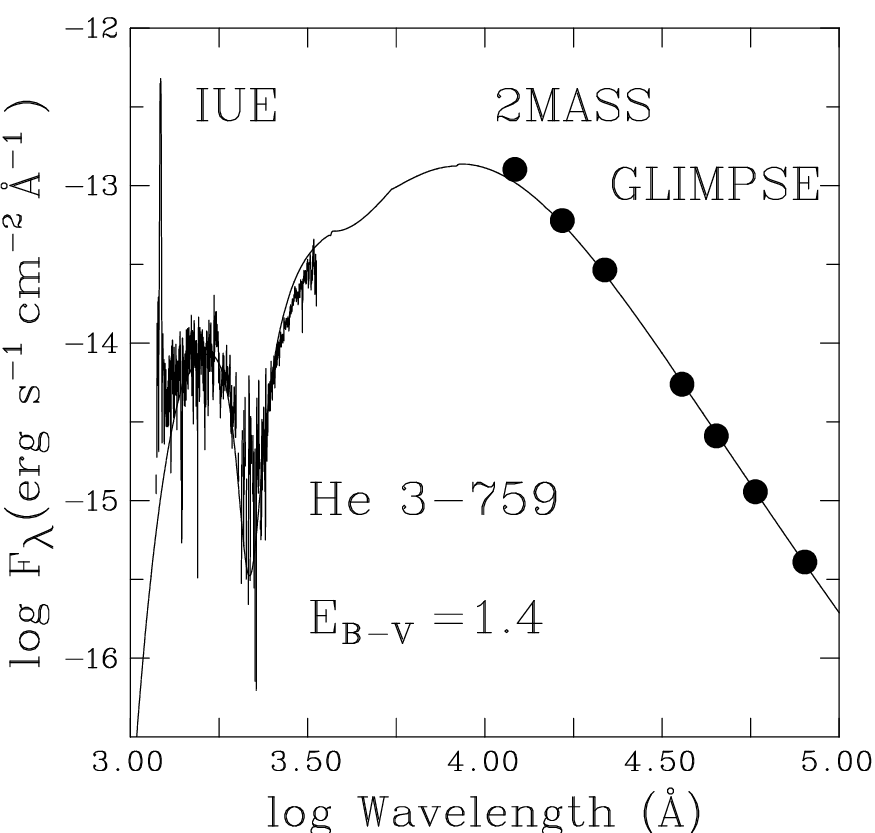}
\caption[]{Reddened model spectral energy distribution of He\,3--759
($E_{B-V}$=1.4, $R_{\rm V}$=3.2) overlaid upon UV (IUE) 
spectrophotometry, plus IR photometry from 2MASS 
\citep{2mass} and GLIMPSE \citep{glimpse}}\label{SED}
\end{center}
\end{figure}

Alternatively, we can exploit archival IUE ultraviolet spectroscopy of
He\,3--759 published by \citet{s90}. We have downloaded low
dispersion, large aperture datasets SWP\,36664 and LWP\,15903 (obtained
on 12 Jul 1989) from the IUE Newly Extracted Spectra
archive\footnote{http://sdc.laeff.inta.es/ines/}.  We have reddened
the spectral energy distribution of our He\,3--759 model from Sect.~\ref{SECT4}
and obtain an optimum fit to the combined UV spectrophotometry and IR photometry
with $E_{B-V}$=1.4 using a standard $R_{\rm V} = A_{\rm V}/E_{B-V}$ = 3.2
extinction law. This is presented in Figure~\ref{SED}, for which overall
agreement is satisfactory, including the comparison with 
Spitzer GLIMPSE \citep{glimpse} at mid-IR wavelengths. Intrinsic colours from our He\,3--759 model
include (K$_{s}$ -- [8.0])$_{0}$ = 0.37$^{\rm m}$ and ([3.6] -- 
[4.5])$_{0}$ = 0.1$^{\rm m}$.

\subsection{Diffuse Interstellar Bands}

We may also exploit the strong DIB features in the visual
spectrum of He\,3--759 with respect to other moderately reddened
stars in Figures~\ref{blue}--\ref{red}, notably  \lam\lam4428, 
5780 and 6613. The DIB at \lam4428 is particularly strong, with an 
equivalent width of 2.5$\pm$0.1\AA, arguing for $E_{B-V}$ in excess
of 1.0 according to \citet{s02}.

Weaker DIB lines are seen to correlate reasonably well with
$E_{B-V}$, in particular \lam\lam5780, 5797 \citep{h93} and \lam8620
\citep{m08}.  Precise measurement of these lines can be complicated, 
e.g., \lam5780 is blended with the broad \lam5778 feature
\citep[cf. Table~A1,][]{h93}.  Continuum placement also introduces 
uncertainties.  Equivalent width ($W_{\lambda}$) estimates for these three 
DIB features
are given in Table~\ref{dibs}, with uncertainties of $\pm$10\% (sufficient
for the purposes of the current investigation). The average of these
three estimates is $E_{B-V}$=1.46$^{\rm m}$. Finally, our FEROS spectroscopy
confirms the claim from \citet{ch79} that the Ca\,{\sc ii} H line is broadened,
albeit owing to stellar H$\epsilon$, rather than being of interstellar 
origin.

\begin{table}[!h]
\caption[]{Equivalent widths ($W_{\lambda}$) of selected diffuse interstellar bands (DIBs) 
and the resulting estimates of $E_{B-V}$.  Uncertainties on the widths
are $\pm$100 m\AA\/ for \lam4428, and $\pm$10\% for the other lines.}\label{dibs}
\begin{center}
\begin{tabular}{cccl}
\hline
Line (\AA) & $W_{\lambda}$ (m\AA) & $E_{B-V}$ & Calibration \\
\hline
4428 &  2500 & $>$1.0 & \citet{s02} \\
5780 & \o705 & 1.38 & \citet{h93} \\
5797 & \o225 & 1.51 & \citet{h93} \\
 8620 & \o550 & 1.50 & \citet{m08} \\
\hline
\end{tabular}
\end{center}
\end{table}

\subsection{Distance to He\,3--759}

The three methods outlined above provide the following estimates
of $A_{\rm V} = R_{\rm V}~E_{B-V}$. IR photometry results in
$A_{\rm V}$ = 5.1 (for $R_{\rm V}$ = 3.1), UV spectrophotometry
implies $A_{\rm V}$ = 4.5 and the line strengths of DIB features also
suggest $A_{\rm V}$ = 4.5, yielding $A_{\rm V} \sim$ 4.7$^{\rm m}$ or 
$A_{\rm K_{s}}$ = 0.53$^{\rm m}$. 
If we assume that He\,3--759 has a similar absolute K$_{s}$-band
magnitude to HD~151804 (O8\,Iaf) we may estimate its distance.

HD~151804 is a member of Sco~OB1 \citep[distance 1.9 kpc, ][]{h78} from 
which $M_{\rm K_{s}}$ = $-$6.7$^{\rm m}$ is obtained (Table~\ref{photom}), giving
a distance modulus of 14.06$\pm$0.5$^{\rm m}$ or
distance of 6.5$^{+1.6}_{-1.3}$ kpc for 
He\,3--759. For an adopted Solar 
galactocentric distance of 8.0 kpc \citep{r93}, He\,3--759 would lie in the 
Sagittarius-Carina arm, close to the Solar circle
$\sim$7.5$^{+0.6}_{-0.4}$ kpc from the Galactic Centre.


\begin{figure}
\begin{center}
\includegraphics[width=\columnwidth]{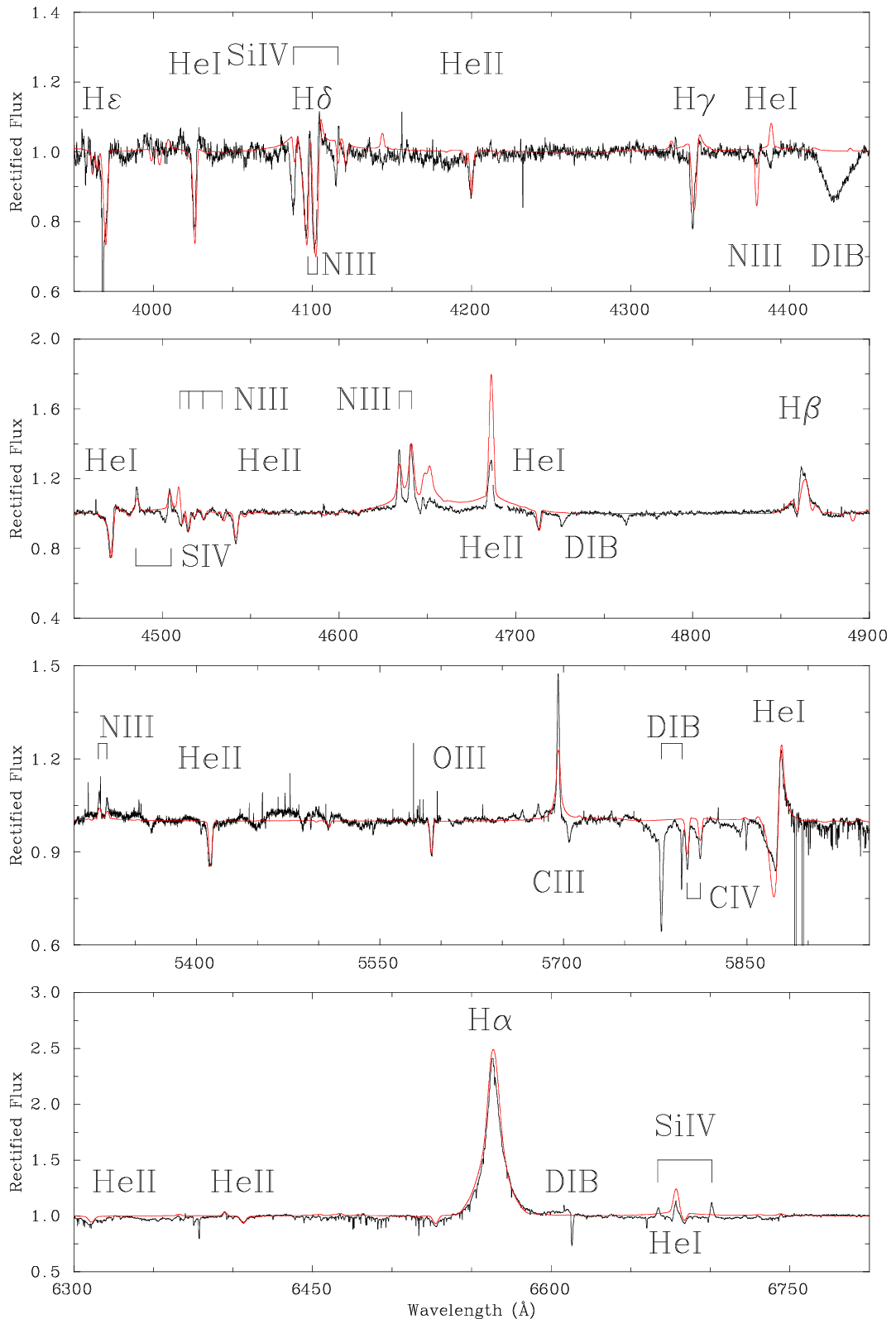}
\caption[]{Spectroscopic fits (dotted, red) to FEROS observations 
(solid, black) of He\,3--759.}\label{fits}
\end{center}
\end{figure}

\section{Physical and Wind Parameters}\label{SECT4}

We have derived the physical and wind properties of He\,3--759
using CMFGEN \citep{cmf}, and re-analysed optical spectroscopy of HD\,151804
(O8\,Iaf) and HD\,152408 (WN9ha) from \citet{cb97} for comparison.

\subsection{Method}

CMFGEN solves the radiative transfer equation in the co-moving frame, 
under the additional constraint of statistical equilibrium. The 
temperature structure is determined by radiative equilibrium. Since CMFGEN 
does not  solve the momentum equation, a density or velocity 
structure is required.  For the supersonic part, the velocity is 
parameterized with a classical $\beta$-type law, with an exponent derived
from fits to H$\alpha$. This is connected to a hydrostatic density structure at 
depth, such that the velocity and velocity  gradient match at the 
interface. The subsonic density structure is set by a corresponding 
$\log g = 3.0$
fully line-blanketed plane-parallel TLUSTY model \citep[v.200, see ][]{lh03}. 
The atomic model is similar to that adopted by \citet{c02}, 
including ions from H, He, C, N, O, Si, P, S and Fe.

\begin{table}[!h]
\caption[]{Physical and wind properties of He\,3--759 with respect
to HD\,151804 and HD\,152408, allowing for an uncertainty in absolute 
magnitude of $\pm$0.5$^{\rm m}$. Clumped mass-loss rates are quoted 
here for volume filling factors of $f$=0.1.}\label{params} \begin{center}
\begin{tabular}{cccl}
\hline
Star & He\,3--759 & HD\,151804 & HD\,152408 \\
     & O8\,Iaf   & O8\,Iaf    & WN9ha \\
\hline
$T_{\rm eff}$ (kK) & 30.5 & 29.0  & 31.8   \\
$R_{\ast}$ ($R_{\odot}$) & 32.1$^{+8.3}_{-6.6}$ & 35.4 & 32.1  \\
$T_{2/3}$ (kK) &  29.3   & 28.1   &  31.3  \\
$\log L/L_{\odot}$ & 5.90$\pm$0.20   & 5.90  & 5.98   \\
$v \sin i$ (km\,s$^{-1}$) & 100 & 104 & 80 \\
$v_{\infty}$ (km\,s$^{-1}$) & 1000 & 1445 & 970 \\
$\dot{M}$ ($M_{\odot}$\,yr$^{-1}$) & 10$^{-5.17\pm0.15}$&
10$^{-5.20}$ & 10$^{-4.94}$ \\
$X_{\rm H}$ (\%) & 49 & 43   & 27  \\
$X_{\rm He}$ (\%) & 49 & 56  &72  \\
$X_{\rm N}$ (\%) &  0.3  & 0.25   & 0.6  \\
$M_{\rm K_{s}}$ (mag) & --6.7$\pm$0.5 & --6.7  & --6.6 \\
\hline
\end{tabular}
\note{Formal uncertainties in $T_{\rm eff}$ are $\pm$0.5 kK, while 
abundances are reliable to within $\pm$5\% (H and He) or a 
factor of two (N)}
\end{center}
\end{table}

We have assumed a depth-independent Doppler profile for all lines when 
solving for the atmospheric structure in the co-moving frame, while in the 
final calculation of the emergent spectrum in the observer's frame, we 
have adopted a uniform turbulence of 50 km\,s$^{-1}$. Incoherent electron 
scattering and Stark broadening for hydrogen and helium lines are adopted. 
Finally, we convolve our synthetic spectrum with a rotational broadening 
profile for which $v \sin i \sim$ 100 km\,s$^{-1}$.
Clumping is incorporated using a volume filling 
factor, $f$, as described in \citet{h03}, with a typical value of $f$=0.1 
resulting in a reduction in mass-loss rate by a factor of $\sqrt{(1/f)} \sim 
3$. 

\subsection{Results for He\,3--759 and other extreme 
supergiants}\label{results}

We derive the stellar temperature of He\,3--759 using diagnostic He\,{\sc 
i} \lam4471, \lam5876, He\,{\sc ii} \lam4542, \lam5411 lines, together 
with H$\alpha$ and H$\beta$ for the mass-loss rate and velocity structure. 
We have estimated a terminal wind velocity of 1000$\pm$300 km\,s$^{-1}$ 
based upon low-resolution IUE observations of C\,{\sc iv} 
$\lambda$1548--51 
using the method of \citet{p94}, while a (slow) velocity law of 
exponent $\beta$=2 is used for  the supersonic velocity structure is used 
since this provides an excellent fit to the H$\alpha$ profile.
Regarding wind clumping in Of supergiants, either He\,{\sc ii} 
\lam4686  or H$\alpha$ are suitable for determination of the volume 
filling factor $f$, if the  velocity law is known.  However, since 
H$\alpha$ is used to estimate the 
velocity law and the peak emission of He\,{\sc ii} \lam4686 is 
very poorly reproduced, an independent determination of $f$ is not 
achievable.

Spectroscopic fits to FEROS observations are presented in 
Figure~\ref{fits}, with a summary of physical and wind 
parameters presented in Table~\ref{params}. 
Overall, the fits are satisfactory, with the exception of He\,{\sc ii}
\lam4686 that is predicted significantly too strongly in emission.
In addition, P Cygni absorption for He\,{\sc i} \lam5876 is also
predicted too strong, and the singlet He\,{\sc i} \lam4143, \lam4387 
lines are predicted to be in emission, yet they are observed in 
absorption. \cite{n06} discuss problems relating to the use of singlet 
He\,{\sc i} lines in O stars, such that triplets (e.g. \lam4471, \lam5876) 
are favoured. We obtain a helium enriched atmosphere with He/H = 0.25 by 
number or $X_{\rm He}$ = 49\% by mass. The prominent N\,{\sc 
iii} \lam\lam4097-4103 and \lam\lam4634-41 features of He\,3--759 are well 
matched using a mass fraction
of $X_{\rm N}$ = 0.3\%, corresponding to an enrichment of 4 times the
solar value. However, N\,{\sc iii} \lam4379 is predicted to be 
too strong and N\,{\sc iii} \lam5320-24 is too weak, such that we admit a
factor of two uncertainty in the nitrogen abundance. Turning to 
other elements, both C\,{\sc iii} \lam5696 and C\,{\sc 
iv} \lam\lam5801-12 favour a high carbon abundance while
\lam\lam4647-51 requires a low abundance. The model presented in 
Fig.~\ref{fits} was obtained for an intermediate abundance of 
$X_{\rm C}$ = 0.2\% (0.7 times the solar case), although large 
uncertainties are admitted. For  oxygen, solely O\,{\sc 
iii} \lam5592 is observed, from which we estimate $X_{\rm O}$ = 
0.2\% (0.5 times the solar value). For silicon, sulphur and iron we adopt 
solar values.

We have also reanalysed two of the reference stars -- HD\,151804 (O8\,Iaf) 
and HD\,152408 (WN9ha) -- based upon our AAT UCLES datasets presented in 
Figs~\ref{blue}--\ref{red} and the method outlined above. A TLUSTY $\log 
g = 3.25$ model at depth was adopted for HD\,152408 since $\log g = 
3.0$ models were not available for $T_{\rm eff}$ = 32.5kK. For current 
stellar masses of $\sim 40 M_{\odot}$ (see Sect.~\ref{evol}), surface 
gravities are 
$\log g \sim 3.0$, while effective gravities, corrected for radiation 
pressure, are $\log g_{\rm eff} \sim 2.8$.\footnote{The Eddington 
parameter -- the ratio of radiation pressure to gravity -- 
is $\Gamma_{e} \sim$ 0.35 for He\,3--759 and HD\,151804.}

Fits are of comparable quality to those presented here for He\,3--759,
also failing to reproduce He\,{\sc ii} \lam4686 emission,
with their physical and wind properties also provided in
Table~\ref{params}. As expected, the physical parameters and chemical 
composition of the three stars are very similar, with the more advanced 
spectral type of WN9ha for HD\,152408 attributable to a somewhat higher 
mass-loss rate - see \citet{bc99} for a general discussion of this 
subject. In addition, the hydrogen contents of HD\,151804 and He\,3--759 
are similar, with a significantly lower hydrogen mass fraction for 
HD\,152408. Subtle differences between the present study and \citet{cb97} 
follow from the improved metal line blanketing (primarily Fe), TLUSTY 
structure at depth and allowance for wind clumping.

\subsection{Comparison with evolutionary
model predictions}\label{evol}

A comparison between the physical properties of He\,3--759 and 
non-rotating, solar metallicity Geneva models from \citet[][see also
\citeauthor{ls01} \citeyear{ls01}]{m94} suggests an age of 2.7 Myr and 
initial mass of $\sim60 M_{\odot}$. Similar results are obtained for 
HD\,151804 and HD\,152408, in good agreement with the age of the NGC~6231 
cluster within Sco~OB1, as derived by \citet{c06} using the same set of 
isochrones. However, these standard evolutionary models are well known not 
to predict the observed helium enrichment at such phases.

In contrast, comparisons with the 
evolutionary models of \citet{mm00} allowing for rotation and
contemporary mass-loss rate prescriptions enable 
reasonable matches to both the  surface hydrogen abundance ($\sim 40$\%) 
and location in the H-R diagram.  For a distance of 6.5 kpc to He\,3--759, 
initial 60\,$M_{\odot}$ models rotating at 300 km\,s$^{-1}$
suggest a greater age of 3.9 Myr, while a slightly lower age of 3.6 Myr is obtained
for a non-rotating 60\,$M_{\odot}$ model. At these ages, current
stellar  masses lie in
the range 35--45 $M_{\odot}$, from which we adopt 40 $M_{\odot}$ for surface gravity
estimates. Lower mass evolutionary  
models from \citet{mm00} fail to predict the combination of surface hydrogen content and 
its position in the H-R diagram, favouring our preferred distance to He\,3--759.

In summary, He\,3--759 appears to be a very high mass star 
at a relatively young age, but unlike HD\,151804 and HD\,152408 it does 
not reside within a known cluster or OB association. According to 
\cite{l03}, the most massive star of a cluster (of mass $M_{\rm clu}$) 
scales with cluster mass according to 1.2 $M_{\rm clu}^{0.45}$ suggesting 
a lower limit of $\sim 6000 M_{\odot}$ for its birth cluster. He\,3--759 
does not possess a high radial velocity so it would be expected to be 
located close to its natal cluster. Alternatively, \citet{p07} have 
proposed that some massive stars may form in relatively low mass clusters.
Such clusters would not necessarily be easily identified at large 
distances, as is the case for He\,3--759.



\section{Summary}\label{SECT5}

We have presented a high quality FEROS spectrum of the poorly
studied, early-type emission line supergiant He\,3--759, from which an 
O8\,Iaf classification is obtained, and clarified its coordinates.
We have used three methods to estimate its high interstellar extinction,
namely fitting a stellar model to its IUE ultraviolet spectrophotometry
and 2MASS and GLIMPSE photometry; obtaining its near-IR extinction from
comparison with intrinsic colours; deriving its visual extinction from
measured strengths of DIBs. Combining these approaches implies
$A_{\rm V}$ = 4.7$^{\rm m}$ or $A_{\rm K_{s}}$ = 0.53$^{\rm m}$.  If we 
assume that 
He\,3--759 has a similar absolute $K_{s}$-band magnitude to HD~151804 
(O8\,Iaf) 
its distance is estimated as 6.5 kpc, within the Sagittarius-Carina arm.
The presence of such a high-mass ($\sim 60M_{\odot}$) star in isolation
is curious given the lack of a nearby cluster, which would be expected to 
be relatively massive ($\geq$6000\,$M_{\odot}$). 

No doubt, many other emission-line OB supergiants await discovery, in
view of large optical surveys such as IPHAS and VPHAS+. 
Alternatively, visibly obscured extreme early-type supergiants may 
be  identified by their infrared free-free excess following the approach
of \citet{h07}. 

\begin{acknowledgements}
We thank John Hillier for his development of CMFGEN, Hugues Sana for 
his reprocessing  of the data, and Martin Cordiner and Keith T. Smith for 
helpful
discussion regarding the interstellar features. This publication is based in part upon INES data from the IUE satellite,
2MASS which is a joint project of the University of Massachusetts and the IPAC/CalTech,  funded by the NASA and the NSF, and
Spitzer datasets from NASA/IPAC Infrared Science Archive (IRSA). IRSA is operated by JPL, CalTech under contract with NASA.
\end{acknowledgements}


\bibliographystyle{aa}

\end{document}